%
% uuencoded figures are supplied in a separate file.
% the following is to incorporate them in the body of the text,
% using epsf.
\let\includefigures=\iftrue
%
% activate this if you don't have epsf.
%\let\includefigures=\iffalse
%
% blackboard bold fonts will be used, if present --
\batchmode
  \font\blackboard=msbm10
\errorstopmode
\newif\ifamsf\amsftrue
\ifx\blackboard\nullfont
  \amsffalse
\fi
\newfam\black
%
% If this presents any difficulties, delete the preceding 8 lines and
% uncomment the next line
%\newif\ifamsf\amsffalse
%
%
\input harvmac.tex
\includefigures
\message{If you do not have epsf.tex (to include figures),}
\message{change the option at the top of the tex file.}
\input epsf
\def\figin{\epsfcheck\figin}\def\figins{\epsfcheck\figins}
\def\epsfcheck{\ifx\epsfbox\UnDeFiNeD
\message{(NO epsf.tex, FIGURES WILL BE IGNORED)}
\gdef\figin##1{\vskip2in}\gdef\figins##1{\hskip.5in}% blank space instead
\else\message{(FIGURES WILL BE INCLUDED)}%
\gdef\figin##1{##1}\gdef\figins##1{##1}\fi}
\def\DefWarn#1{}
\def\figinsert{\goodbreak\midinsert}
\def\ifig#1#2#3{\DefWarn#1\xdef#1{fig.~\the\figno}
\writedef{#1\leftbracket fig.\noexpand~\the\figno}%
\figinsert\figin{\centerline{#3}}\medskip\centerline{\vbox{\baselineskip12pt
\advance\hsize by -1truein\noindent\footnotefont{\bf Fig.~\the\figno:} #2}}
\bigskip\endinsert\global\advance\figno by1}
%%%
\else
\def\ifig#1#2#3{\xdef#1{fig.~\the\figno}
\writedef{#1\leftbracket fig.\noexpand~\the\figno}%
%\figinsert\figin{\centerline{#3}}\medskip\centerline{\vbox{\baselineskip12pt
%\advance\hsize by -1truein\noindent\footnotefont{\bf Fig.~\the\figno:} #2}}
%\bigskip\endinsert
\global\advance\figno by1}
\fi
\ifamsf
%\font\blackboard=msbm10
\font\blackboards=msbm7
\font\blackboardss=msbm5
\textfont\black=\blackboard
\scriptfont\black=\blackboards
\scriptscriptfont\black=\blackboardss
\def\Bbb#1{{\fam\black\relax#1}}
\else
\def\Bbb{\bf}
\fi
% *************************************
%\draft
%
\def\yboxit#1#2{\vbox{\hrule height #1 \hbox{\vrule width #1
\vbox{#2}\vrule width #1 }\hrule height #1 }}
\def\fillbox#1{\hbox to #1{\vbox to #1{\vfil}\hfil}}
\def\ybox{{\lower 1.3pt \yboxit{0.4pt}{\fillbox{8pt}}\hskip-0.2pt}}

\def\comments#1{}

\def\p{\partial}

\def\half{{1\over 2}}
\def\Tr{{{\rm Tr\  }}}

\def\CD{{\cal D}}

\def\CF{{\cal F}}

\def\CM{{\cal M}}
\def\CN{{\cal N}}

\def\ap{\alpha'}

\def\I{I}

\def\II{\relax{I\kern-.10em I}}
\def\IIa{{\II}a}
\def\IIb{{\II}b}

\def\cascade{{\cal A}}

\def\IZ{\relax\ifmmode\mathchoice
{\hbox{\cmss Z\kern-.4em Z}}{\hbox{\cmss Z\kern-.4em Z}}
{\lower.9pt\hbox{\cmsss Z\kern-.4em Z}}
{\lower1.2pt\hbox{\cmsss Z\kern-.4em Z}}\else{\cmss Z\kern-.4em
Z}\fi}
\def\IB{\relax{\rm I\kern-.18em B}}
\def\IC{{\relax\hbox{$\inbar\kern-.3em{\rm C}$}}}
\def\ID{\relax{\rm I\kern-.18em D}}
\def\IE{\relax{\rm I\kern-.18em E}}
\def\IF{\relax{\rm I\kern-.18em F}}
\def\IG{\relax\hbox{$\inbar\kern-.3em{\rm G}$}}
\def\IGa{\relax\hbox{${\rm I}\kern-.18em\Gamma$}}
\def\IH{\relax{\rm I\kern-.18em H}}
\def\II{\relax{\rm I\kern-.18em I}}
\def\IK{\relax{\rm I\kern-.18em K}}
\def\IP{\relax{\rm I\kern-.18em P}}
%\def\IX{\relax{\rm X\kern-.01em X}}
%this doesn't work

%

\def\Hom{{\rm Hom}}
\def\inbar{\,\vrule height1.5ex width.4pt depth0pt}

\def\p{\partial}

\font\cmss=cmss10 \font\cmsss=cmss10 at 7pt
\def\IR{\relax{\rm I\kern-.18em R}}

\def\Tr{\rm Tr}
\def\ker{{\rm ker\ }}

\def\Tr{\mathop{\rm Tr}\nolimits}
\def\th{^{\rm\scriptstyle th}}
\def\gitquot{\mathchoice
{\mathrel{\mskip-4.5mu/\!/\mskip-4.5mu}}
{\mathrel{\mskip-4.5mu/\!/\mskip-4.5mu}}
{\mathrel{\mskip-3mu/\mskip-4.5mu/\mskip-3mu}}
{\mathrel{\mskip-3mu/\mskip-4.5mu/\mskip-3mu}}}
\ifamsf
\def\IC{\Bbb{C}}
\def\IP{\Bbb{P}}
\def\IR{\Bbb{R}}
\def\IZ{\Bbb{Z}}
\fi
\def\BR{\IR}
\def\BZ{\IZ}
\def\BP{\IP}
\def\BC{\IC}

\def\lp10{l_P^{10}}
\def\lp11{l_P^{11}}
\def\R11{R_{11}}

\newbox\tmpbox\setbox\tmpbox\hbox{\abstractfont RU-97-12, CU-TP-823,
IASSNS-HEP-97/24}
\Title{\vbox{\baselineskip12pt\hbox to\wd\tmpbox{\hss
hep-th/9704151}\hbox{RU-97-12, CU-TP-823, IASSNS-HEP-97/24}}}
{\vbox{
\centerline{Orbifold Resolution by D-Branes} }}
\centerline{Michael R. Douglas,$^1$ Brian R. Greene,$^{2,}$%
\footnote{${}^{*}$}{On leave from Department of
Physics, Cornell University, Ithaca, NY 14850.} and %
David R. Morrison$^{3,}$\footnote{${}^{\dagger}$}{On leave from Department of
Mathematics, Duke University, Durham, NC 27708-0320.}}
\medskip
\centerline{$^1$Department of Physics and Astronomy}
\centerline{Rutgers University }
\centerline{Piscataway, NJ 08855--0849}
\centerline{\tt mrd@physics.rutgers.edu}
\medskip
\centerline{$^2$Departments of Physics and Mathematics}
\centerline{Columbia University }
\centerline{New York, NY 10025}
\centerline{\tt greene@math.columbia.edu}
\medskip
\centerline{$^3$Schools of Mathematics and Natural Sciences}
\centerline{Institute for Advanced Study}
\centerline{Princeton, NJ 08540}
\centerline{\tt drm@math.duke.edu}
\medskip
\bigskip
\noindent
We study topological properties of the D-brane resolution of
three-dimensional orbifold singularities, $\BC^3/\Gamma$, for finite
abelian groups $\Gamma$. The D-brane vacuum moduli space is shown to fill
out the background spacetime with Fayet--Iliopoulos parameters controlling
the size of the blow-ups. This D-brane vacuum moduli space can be
classically described by a gauged linear sigma model, which is shown to be
non-generic in a manner that projects out non-geometric regions in its
phase diagram, as anticipated from a number of perspectives.

\Date{April 1997}
%\draft
\lref\bss{T. Banks, N. Seiberg, and E. Silverstein, ``Zero and
One-dimensional Probes with $N{=}8$ Supersymmetry,'' hep-th/9703052.}
\lref\DHVW{L. Dixon, J. A. Harvey, C. Vafa, and E. Witten, ``Strings on
Orbifolds, I, II'' Nucl. Phys. B261 (1985) 678; Nucl. Phys. B274 (1986) 285.}
\lref\itoreid{Y. Ito and M. Reid, ``The McKay Correspondence for Finite
Subgroups of SL(3,\BC),'' in: {\it Higher Dimensional Complex Varieties}\/
(M.~Andreatta et al., eds.), de Gruyter, 1996, p.~221; alg-geom/9411010.}
\lref\reid{M. Reid, ``McKay Correspondence,'' alg-geom/9702016.}
\lref\GanMorSei{O. J. Ganor, D. R. Morrison, and N. Seiberg, ``Branes,
Calabi--Yau Spaces, and Toroidal Compactification of the $N{=}1$
Six-Dimensional $E_8$ Theory,'' Nucl. Phys. B487 (1997) 93;
hep-th/9610251.}
\lref\MPorb{D. R. Morrison and M. R. Plesser, to appear.}
\lref\sag{A. Sagnotti, ``Some Properties of Open-String Theories,''
hep-th/9509080.}
\lref\kron{P. B. Kronheimer, ``The Construction of ALE Spaces as
Hyper-K\"{a}hler Quotients,'' J. Diff. Geom.  29 (1989) 665.}
\lref\infirri{A. V. Sardo Infirri, ``Partial Resolutions of Orbifold
Singularities via Moduli Spaces of HYM-type Bundles,'' alg-geom/9610004.}
\lref\infirritwo{A. V. Sardo Infirri, ``Resolutions of Orbifold Singularities
and Flows on the McKay Quiver,'' alg-geom/9610005.}
\lref\polcai{J.~Polchinski and Y.~Cai, ``Consistency of Open Superstring
Theories,'' Nucl. Phys.  B296 (1988) 91.}
\lref\bwb{M. R. Douglas, ``Branes within Branes,'' hep-th/9512077.}
\lref\dm{M. R. Douglas and G. Moore, ``D-Branes, Quivers, and ALE Instantons,''
hep-th/9603167.}
\lref\dg{M. R. Douglas and B. R. Greene, to appear.}
\lref\dta{M. R. Douglas, to appear.}
\lref\JM{C. Johnson and R. Myers, ``Aspects of Type IIB Theory on ALE
Spaces,'' hep-th/9610140.}
\lref\egs{M. R. Douglas, ``Enhanced Gauge Symmetry in M(atrix) Theory,''
hep-th/9612126.}
\lref\polpro{J.~Polchinski, ``Tensors from K3 Orientifolds,''
hep-th/9606165.}
\lref\BFSS{T. Banks, W. Fischler, S. H. Shenker and L. Susskind,
``M Theory as a Matrix Model: A Conjecture,'' hep-th/9610043.}
\lref\ooy{H. Ooguri, Y. Oz and Z. Yin, ``D-Branes on Calabi--Yau Spaces and
Their Mirrors,'' Nucl.Phys. B477 (1996) 407; hep-th/9606112.}
\lref\agm{P. S. Aspinwall, B. R. Greene and D. R. Morrison, ``Calabi--Yau
Moduli Space, Mirror Manifolds and Spacetime Topology Change in
String Theory,'' Nucl. Phys. B416 (1994) 414; hep-th/9309097.}
\lref\rAGMsd{P. S. Aspinwall, B. R. Greene and D. R. Morrison, ``Measuring
Small Distances in $N{=}2$ Sigma Models,''
Nucl. Phys. B420 (1994) 184; hep-th/9311042.}
\lref\aspinwall{P. S. Aspinwall, ``Enhanced Gauge Symmetries and K3
Surfaces,'' Phys. Lett. B357 (1995) 329; hep-th/9507012.}
\lref\dos{M. R. Douglas, H. Ooguri and S. H. Shenker, ``Issues in M(atrix)
Theory Compactification,'' hep-th/9702203.}
\lref\rWP{E. Witten, ``Phases of $N{=}2$ Theories In Two Dimensions,''
Nucl. Phys. B403 (1993) 159; hep-th/9301042.}
\lref\witPT{E. Witten, ``Phase Transitions In M-Theory And F-Theory,''
Nucl. Phys. B471 (1996) 195; hep-th/9603150.}
\lref\fulton{W. Fulton, {\it Introduction to Toric Varieties,}
Princeton University Press, 1993.}
\lref\oda{T. Oda, {\it Convex Bodies and Algebraic Geometry,}
Springer-Verlag, 1988.}
\lref\rGK{B. R. Greene and Y. Kanter, ``Small Volumes in Compactified
String Theory,'' hep-th/9612181.}
\lref\rAG{P. S. Aspinwall and B. R. Greene, ``On the Geometric
Interpretation of $N{=}2$ Superconformal Theories,''
Nucl. Phys. B437 (1995) 205; hep-th/9409110.}
\lref\mp{D. R. Morrison and M. R. Plesser, ``Summing the Instantons:
Quantum Cohomology and Mirror Symmetry in Toric
Varieties,'' Nucl. Phys. B440 (1995) 279; hep-th/9412236.}
\lref\rDelzant{T.~Delzant, ``{H}amiltoniens p\'eriodiques et images convexe de
  l'application moment,'' Bull. Soc. Math. France {\bf 116} (1988) 315.}
\lref\rAudin{M.~Audin, {\it The Topology of Torus Actions on Symplectic
Manifolds}, Birkh\"auser, 1991.}
\lref\rCox{D. A. Cox, ``The Homogeneous Coordinate Ring of a Toric
Variety,'' J. Algebraic Geom. 4 (1995) 17; alg-geom/9210008.}
\lref\gp{E. G. Gimon and J. Polchinski,
``Consistency Conditions for Orientifolds and D-Manifolds,'' Phys. Rev. D54
(1996) 1667; hep-th/9601038.}
%
% forward equation references
%
\newsec{Introduction}

D-branes are interesting probes of topology and geometry---as explored in
many recent works,  spacetime is a derived concept from the D-brane
perspective, emerging
from non-trivial moduli spaces of D-brane world-volume gauge theories.

Orbifold resolution was the first non-trivial example of this phenomenon,
studied in \refs{\dm,\polpro,\JM}.
The classical Lagrangian of the probe theory is a projection of maximally
supersymmetric
super Yang--Mills theory, with Fayet--Iliopoulos terms controlled by
twist sector moduli.  In the two dimensional
case--- $\BC^2/\Gamma$ ---it was shown that the classical gauge theory
moduli space is
the resolved orbifold,  and the metric is computable.
The result is a physical realization of the hyper-K\"ahler quotient
construction of Kronheimer \kron.

Here we demonstrate that an analogous procedure works for
cyclic orbifold singularities in three complex dimensions, treating
the simplest examples $\BC^3/\BZ_3$ and $\BC^3/\BZ_5$ explicitly.
The world-volume theory has the equivalent of
$d=4$, $\CN=1$ supersymmetry, and
the superpotential and D-terms are no longer related by supersymmetry.
The mathematical counterpart of this statement is that the construction
is not a hyper-K\"ahler quotient but rather a blow-up of a singular K\"ahler
quotient.  The superpotential defines the complex structure and the blow-up,
while the D-terms define the periods of the metric (though the full metric
depends on both data).
The resulting theories are physical realizations of a generalization
of Kronheimer's construction to $\BC^n/\Gamma$ orbifold resolution,
recently developed by Sardo Infirri \refs{\infirri,\infirritwo}.

In this note we discuss topological and qualitative aspects of the
construction.  We show that a generic deformation of the moduli from
the orbifold point leads to complete resolution of the orbifold and
produces a smooth metric.

At weak string coupling, where our analysis is directly applicable,
D-branes provide a description of spacetime at
short distances complementary to the closed string approach of
\refs{\agm,\rWP,\rAGMsd}.
In that approach, it was found that the K\"ahler moduli spaces of sigma model
compactification form but a small subset of the possibilities.
Linear sigma models---when available---provide a more general definition
and reveal a rich
phase structure of the moduli space including target spaces of different
topology and non-geometric phases.
Computations using mirror symmetry
established that all of these phases are connected, and gave
physical results suggesting a more geometrical picture even for the
non-geometric phases.

By using D-brane gauge theory,
we can study topological properties of orbifolds on length scales $r$
in the range $ l_p^{11} < r  < \sqrt \alpha'$. The classical physics of
these probe theories turns out to also be describable by
gauged linear sigma models,
and {\it a priori}, the moduli space of these linear sigma models could
again have numerous phases.
We show, though, that the D-brane linear sigma models
are not generic in that they only probe part
of the linear sigma model moduli space. We present evidence
that this subspace is nothing but the ``partially enlarged K\"ahler
moduli space,'' every point of which has a geometrical interpretation.
Thus D-branes appear to directly  project out the non-geometric phases.

We believe that this D$0$-brane description remains qualitatively correct even
at strong string coupling, following the lines of \refs{\BFSS,\egs,\dos}.
(This issue will be discussed further in \dg.)
If so, our results confirm
an idea of Witten that in M theory only geometric
phases appear \witPT.

In section 2, we review some relevant aspects
of the stringy description of the moduli
space of Calabi--Yau compactification.
In section 3 we discuss the D-brane construction.
In section 4 we derive some of the qualitative features the moduli spaces,
using the gauged linear sigma model description.
Section 5 gives conclusions and some open questions.

\newsec{Quantum geometry of Calabi--Yau manifolds}

At present there are two limits near which one can understand
the quantum moduli space of K\"ahler structures
on a Calabi--Yau manifold: weakly coupled \IIa\ (or \IIb) string theory and
M theory.
As the explicit analysis we carry out is justified in the weak string
coupling limit, we first describe our work in the \IIa\
framework;  subsequently, we comment on its possible
relevance in M theory, assuming unexpected surprises do not
arise when extrapolating to strong coupling.

The starting point is a classical
moduli space of metrics on a Calabi--Yau manifold $M$.  This
is locally a product of a space of complex structures $\CM_C$ and
complexified K\"ahler forms $\CM_K$.
Classically, $\CM_K$ is a complexification of the K\"ahler cone $\CM_{Kc}$,
the subset of $H^2(M,\BR)$
for which the volumes of cycles are positive.
As such, it appears to have boundaries, each of which requires
physical explanation.
{\it A priori}\/ these explanations would be expected to involve quantum
effects, and this is the subject of quantum geometry.

For our present purpose
it is worthwhile to build up to
the quantum corrected version of $\CM_{Kc}$ in
two steps.
First, one must augment the complexification of $\CM_{Kc}$ to
$\CM_{Kc}^{\rm p.e.}$,
the {\it partially enlarged K\"ahler moduli space}, which consists
of the classical complexified K\"ahler cones of all Calabi--Yau manifolds
related to $M$ by flops along rational curves, glued along common faces.
Second,  realizing that this does not exhaust the full moduli space
\refs{\agm,\rWP},
one must augment  $\CM_{Kc}^{\rm p.e.}$
by gluing on additional regions (or ``phases'')
which arise when higher dimensional
subspaces of $M$ or its birational cousins are shrunk to points.

Quantum corrections can also change the boundaries between
different phases.  In weakly coupled string theory, these changes are by
effects
of order $\ap$.  Thus it may be that even in a phase in which a cycle
shrinks to a point, its observed volume is non-zero.

The final result is sometimes called
the {\it enlarged nonlinear sigma model K\"ahler moduli space}\/
$\CM_{Kc}^{\rm e.n.}$, though it is already a good description
of $\CM_K$ at weak string coupling.

There are two approaches to carrying out the second step.
One can
seek a mirror manifold $W$ of $M$ and study its complex structure
moduli space using standard techniques from algebraic geometry,
or one can seek a gauged linear sigma model description of $M$.
Each has its advantages and disadvantages. The linear sigma
model approach allows us to physically identify all of the additional
regions in  $\CM_{Kc}^{\rm e.n.}$.
These are usually called ``non-geometrical phases'' since
they are not interpretable in terms of smooth Calabi--Yau sigma models.
Instead, they are most naturally described in terms of more abstract
field theories such as orbifolds, Landau--Ginsburg models, and various
hybrid combinations.  However,  unlike the mirror symmetry approach,
the linear sigma model does not construct  $\CM_{Kc}^{\rm e.n.}$
directly. Rather,  it builds an auxiliary moduli space---the {\it enlarged
linear sigma model moduli space}\/  $\CM_{Kc}^{\rm e.l.}$.
The latter is discussed in detail in \rWP\ as well as
\refs{\agm,\rAGMsd,\rAG,\mp}; it
fills out all of $H^2(M,\IC)$, for example.  The space $\CM_{Kc}^{\rm e.l.}$
 must then be subjected to  a renormalization
group  action in order to yield the moduli
space $\CM_{Kc}^{\rm e.n.}$ for conformally invariant physical models.
In other words, the
linear sigma model parameters mapped out by $\CM_{Kc}^{\rm e.l.}$
are secondary constructs which can be used to label
universality classes of physical string models; their primary physical
counterparts,  however, are the
parameters which arise upon passing to the conformal limit.

Of course, regardless of which approach one follows, the final form for
$\CM_{Kc}^{\rm e.n.}$ is the same.
The essential properties of $\CM_{Kc}^{\rm e.n.}$
 for our present concerns are three-fold. First,
$\CM_{Kc}^{\rm e.n.}$ has numerous phase regions, some of which
have a manifest geometric interpretation in terms of Calabi--Yau nonlinear
sigma models, while others do not.
The boundaries in  $\CM_{Kc}^{\rm e.n.}$  are not
sharp phase transitions but rather delineate the edges of perturbative
convergence in each region. One can pass from region to region
by analytic continuation.
 Second, the boundaries
of the classical geometrical moduli space $\CM_{Kc}^{\rm p.e.}$
get shifted by terms of order $\alpha'$ in the quantum corrected
moduli space $\CM_{Kc}^{\rm e.n.}$.
Third,  $\CM_{Kc}^{\rm e.n.}$
appears to be a subset of the classical moduli space
$\CM_{Kc}^{\rm p.e.}$.
This inclusion has not
been proven, but strong circumstantial evidence has been
presented in \rAGMsd. Assuming it to be true, we learn that
every point in $\CM_{Kc}^{\rm e.n.}$ either has a geometrical or
analytically continued geometrical interpretation.
For instance, a Landau--Ginsburg, hybrid or orbifold phase can
be interpreted as the analytic continuation of a smooth Calabi--Yau
sigma model, beyond the realm of sigma model perturbation theory, (i.e.,
to distance scales on the order of $\sqrt {\alpha'}$).

As D-branes are natural probes of sub-stringy distance scales,
it is important to understand the geometrical structure they
sense at such points in moduli space.  D-brane geometry
arises in a different manner from the traditional fundamental
string approach. Whereas a classical background geometry
is generally introduced in fundamental string theory and then
subjected to various corrections, D-brane geometry emerges
from the study of the D-brane vacuum moduli space.
In general this space also undergoes quantum corrections, but in
the particular case of D$0$-branes at weak string
coupling, these are suppressed.
Thus the geometry can be accessed through a classical study of
the vacuum structure of a world volume super-Yang--Mills theory.

In our present study of orbifold resolution,
this world-volume theory will turn out to be classically describable as
a gauged linear sigma model, which would seem to fit well
with the fundamental string description above.
However, closer inspection reveals a crucial difference:
the linear sigma model describing D-branes
is not the one usually used to study fundamental strings.
Now in the fundamental string context, it did not really matter which
linear sigma model was used as a starting point.  Only properties
of the conformally invariant limit were observable, and
a central ingredient in this analysis was the associated use of the
renormalization group to pass to the conformally invariant limit.
It was this
process that gave an analytically continued geometrical interpretation
to the ``non-geometric'' phases. For D-branes there is no
analogous step. If the D-brane linear sigma model analysis were to yield
non-geometric phases, we would have to live with them.

In the following sections we shall see that
the D-brane linear sigma model differs
from that arising from fundamental strings in just the right way to
yield {\it a compatible spacetime interpretation while avoiding the
non-geometric phases}.   At least in the orbifold examples we study,
spacetime at short distances as probed by D-branes agrees with
the analytically continued geometry probed by fundamental strings.
In a novel fashion, D-branes avoid the non-geometrical phases.

What happens if we try to extrapolate this picture to large string coupling?
In the M theory limit, the string scale is far below the
eleven-dimensional Planck scale, leading to the conjecture that
additional quantum effects vanish \witPT.
Specifically, in \witPT\ it was argued that the non-geometric phases
in the type \IIa\ vector multiplet moduli
space are squeezed to zero size. What this means is that
the geometrical phases contained in
$\CM_{Kc}^{\rm e.n.}$ ---the conformal
limit of  $\CM_{Kc}^{\rm p.e.}$ ---expand to fill out the region delineated
by the
classical counterpart,  $\CM_{Kc}^{\rm p.e.}$,
while the non-geometrical phases are flattened out to constitute
the boundaries. In other words, all physical models again have
a geometrical interpretation, except now we do not even have to perform
analytic continuation to find it. This is precisely in accord with
our D-brane picture which avoids  non-geometrical phases from
the word go. In this sense, the strong coupling extrapolation of
our analysis confirms part of the M-theory picture of \witPT.
It would very interesting to extend this analysis to
the other ``non-geometrical'' phases such as Landau--Ginsburg models
and hybrids and see if the same conclusion follows.

An important conclusion of \rAGMsd\ is that  conformal field theory
orbifold points have a nonzero $B$-field. This implies, as pointed out
in \aspinwall, that wrapped D$2$-branes and D$4$-branes have nonzero
mass at the deep interior point of an orbifold phase. On the other hand,
there are boundary points of the orbifold phase where the quantum volume
of cycles associated with the singularity do vanish, and it is here that
wrapped D-branes become massless. In the strong string coupling
limit,  as the orbifold phase gets squeezed, the conformal field theory
orbifold point and the singular boundary point coincide.\foot{This can also
seen by analyzing the orbifold point directly in M theory
compactified on the product of $S^1$ with a three-dimensional orbifold
\refs{\GanMorSei,\MPorb}.}
 The wrapped
D-branes at the non-singular orbifold point should become massless
in this limit. We can
explicitly see this in our D-brane analysis by identifying wrapped
states with ``fractional'' D-brane configurations. We will show that
the masses of these states
go to zero as $1/g_s$ in the strong string coupling limit.

\newsec{$\BC^3/\Gamma$ orbifold compactification}

We consider a type \II\ string theory compactified
on $\BC^3/\Gamma$, with $\Gamma$ a $\BZ_n$ subgroup
of $SU(3)$.
We take $n$ odd to get an isolated singularity.
It is well known
that these orbifolds can be resolved to smooth
spaces with $b_3=0$; in the simplest example of $\BC^3/\BZ_3$ this is
just a blow-up replacing the singularity with a $\BP^2$.
As it happens, in the general case $b_2 = b_4 = (n-1)/2$.

We take complex coordinates $Z^i$ and a generator $g$ of $\BZ_n$
labeled by three integers $(a_1,a_2,a_3)$ with $a_1+a_2+a_3\equiv
0 \, (\hbox{mod } n)$.
It acts as
$Z^i\rightarrow \omega^{a_i} Z^i$ with $\omega\equiv \exp 2\pi i/n$.
Thus the unbroken $d=4$ supersymmetry will be $\CN=2$ in the
closed string sector, and $\CN=1$ on the D-branes.

\subsec{Closed string spectrum}

The closed string sector has $n-1$ twist sectors.
In type \II\ theory each contains a
complex NS-NS field $\phi_k$,
and a complex R-R field making up a hyper (\IIb)
or vector (\IIa) multiplet.
There is a reality condition \DHVW\
\eqn\twreal{\phi_{n-k} = \phi_k^*}
leaving $(n-1)/2$ complex NS-NS fields.
These are the fields which in the large volume limit
become the complexified K\"ahler forms $B+iJ$ on the resolution.
The R-R fields in the $k\th$ twisted sector are a complex bispinor in $d=4$
satisfying the same reality condition.
These decompose into complex $p$-form potentials
$\tilde C^{(p)}_k$,
where the GSO projection
enforces two properties (see for example \refs{\polcai,\bwb}): first, $p$
must be even for \IIb\ and odd for \IIa,
and second, $d\tilde C^{(p)}_k=i*d\tilde C^{(2-p)}_{k}$.
These $p$-form potentials correspond in the large volume limit to
integrals of the bulk R-R fields over various cycles.
In \IIb, these are $\int_{\Sigma_2} C^{(2)}$ and $\int_{\Sigma_4} C^{(4+)}$
(since the fields $\int_{\Sigma_2} C^{(4+)}$ are redundant due to the
self-duality
of $C^{(4+)}$), while in \IIa, they are $\int_{\Sigma_2} C^{(3)}$
(whose duals are of the form $\int_{\Sigma_4} C^{(5)}$).

The type \I\ spectrum is the subsector of type \IIb\
surviving the $\Omega$ (world-sheet twist) projection.
This removes the $B$ moduli, leaving $k$
NS-NS moduli to control the real K\"ahler forms, and $k$ R-R scalars.

\subsec{D-brane theory}

On physical grounds, we expect type \II\ compactification on
the large volume limit of a Calabi--Yau to contain the same D$p$-branes
as the uncompactified theory, with D-branes
wrapped about supersymmetric cycles giving BPS states.

As curvatures approach the string scale, stringy corrections to the
world-volume Lagrangians of these branes could become important.
For example, a single D$0$-brane will have a world-volume kinetic
term incorporating a metric $g_{\mu\nu}$ which, although Ricci-flat
in the small curvature limit, {\it a priori}\/ need not be, just as the
superstring
sigma model metric satisfies a modified Einstein equation with corrections
of order $O(\ap^3)$.  The D$0$-brane metric need not be exactly equal
to the string metric and at this writing
explicit results for these metrics have not been obtained.
However it seems likely that such
corrections are present.

Conformal field theory provides a more general language for compactification,
and D-branes in this general context have been discussed in \ooy.
It is a very interesting question whether moduli spaces of superconformal
boundary conditions are always
similar to the geometric moduli spaces of the
sigma model D-branes.

Here we discuss D-branes defined on an orbifold and related theories
obtained by adding twisted sector moduli with small coefficients,
following the approach of \dm\ in two complex dimensions.
The basic results we rely on are, first,
an analysis of world-sheet consistency conditions which shows that the
orbifold prescription of closed string theory also applies to open string
theory, with all new ``open string twisted sectors'' provided by adding
images of the original D-branes.\foot{The essential features of this
analysis can be found in \gp\ and \dm;  related issues are treated
in \sag\ and references therein. We also thank T. Banks
and M. Berkooz for discussions on this topic.}
Second, analyticity of conformal perturbation
theory around the orbifold point guarantees that the D-brane world-volume
Lagrangian will be the projected flat space Lagrangian with corrections
analytic in the blowup parameters.

Thus a Dirichlet $p$-brane at a point in $\BC^3/\Gamma$
will be defined as the quotient of the theory of
$n$ Dirichlet $p$-branes in $\BC^3$ by a combined
action of $\Gamma$ on $\BC^3$ and the D-brane index (Chan--Paton factor).
A single Dirichlet $p$-brane is described by taking the Chan--Paton factors in
the
regular representation of $\Gamma$.
It is also possible to consider other representations as in
\egs\ (intuitively, leaving out some of the images)
as we discuss below.

The D-brane theory on $\BC^3$ is dimensionally reduced
$d=4$, $\CN=4$ supersymmetric gauge theory with gauge group $U(n)$.
The generator $g$ now acts as above on $\BC^3$ and
on the Chan--Paton factors as the regular representation,
$\gamma(g)_{ij} = \delta_{ij} \omega^i$.
The gauge field projection is
\eqn\orbgpro{
A_{i,j} = \omega^{i-j} A_{i,j}
}
leaving the subgroup $U(1)^n$ unbroken.

The complex positions are the bosonic components of chiral multiplets.
They must satisfy
\eqn\orbpro{
X^\mu_{i,j} = \omega^{i-j+a_\mu} X^\mu_{i,j}.
}
Thus, the non-zero fields take the form $X^\mu_{i,i+a_\mu}$, and each such
field is charged under two $U(1)$'s.
All fields are uncharged under the diagonal $U(1)$.

The action on the fermions is determined the same way, but now taking
$g$ to act on spinors.  For $\Gamma\subset SU(3)$, the $\CN=4$ gaugino will
decompose in the $\CN=1$ theory into a singlet (partners of \orbgpro) and
triplet (partners of \orbpro).
For $\Gamma\not\subset SU(3)$, the projection will generally lead to a
non-supersymmetric theory.

This spectrum can be described by a quiver diagram as in \dm, with a node
corresponding to each $U(1)$ factor of the gauge group, and an oriented
link for each chiral multiplet (as in figure 1).
\ifig\figone{Quiver diagram for $n=5$ and $\vec a=(1,1,3)$.}
{\epsfxsize2.0in
\epsfbox{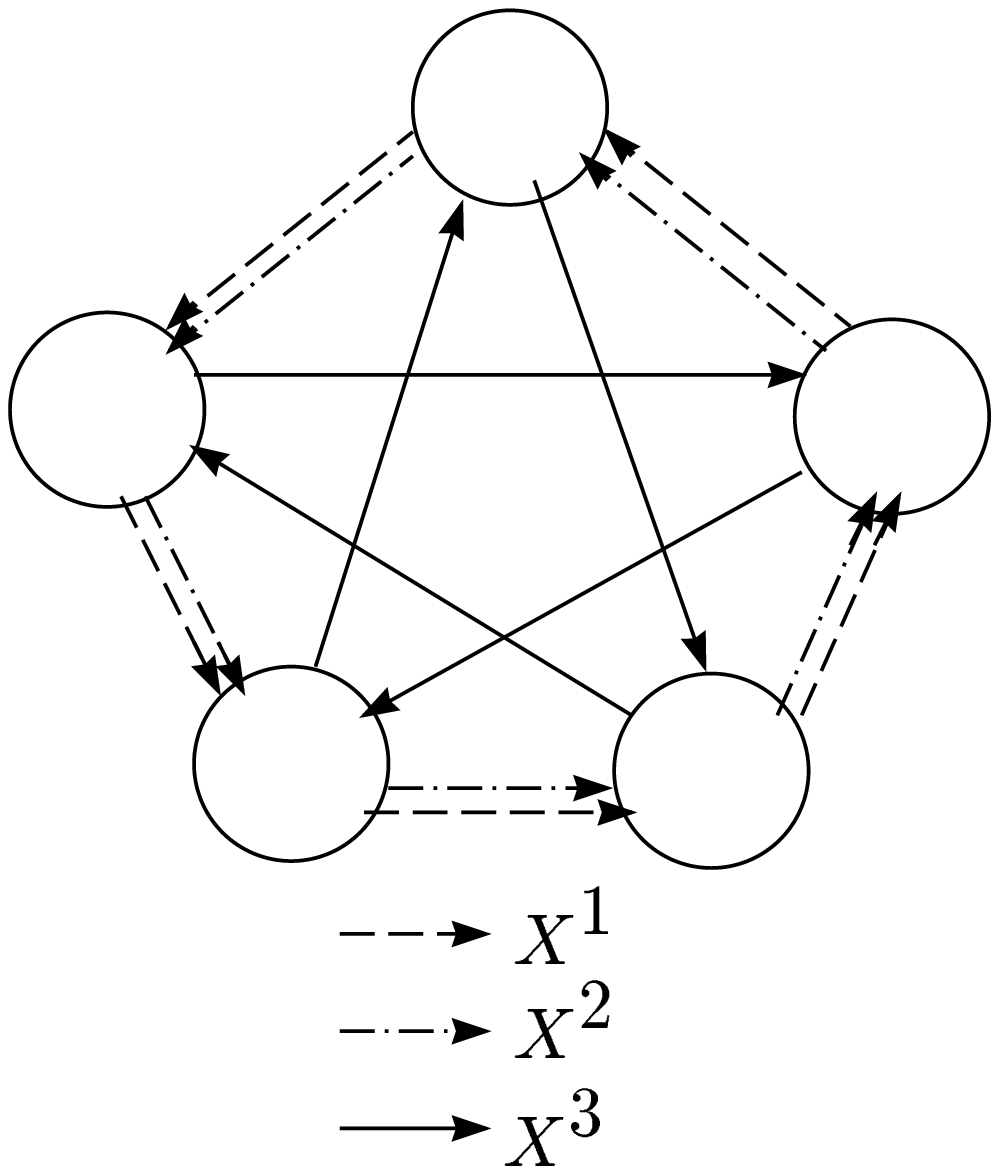}}

The Lagrangian is simply the $\CN=4$ Lagrangian evaluated on fields
satisfying the projection, including a Fayet--Iliopoulos term with
coefficient $\zeta_i$ associated to the $i\th$ $U(1)$, for each $i$.
Supersymmetric vacua will satisfy D-flatness conditions,
\eqn\dterms{
\zeta_i = -\sum_{\mu=1}^3 |X^\mu_{i,i+a_\mu}|^2 + |X^\mu_{i-a_\mu,i}|^2,
}
with $\sum \zeta_i=0$ required for supersymmetric vacua to exist.

The superpotential is the reduction of the $\CN=4$ superpotential,
\eqn\superp{\eqalign{
W &= \Tr [X^1,X^2]X^3 \cr
&= \sum_i X^1_{i,i+a_1} X^2_{i+a_1,i+a_1+a_2} X^3_{i+a_1+a_2,i} -
X^2_{i,i+a_2} X^1_{i+a_2,i+a_2+a_1} X^3_{i+a_2+a_1,i}.
}}
The vacuum satisfies $\p W/\p X^\mu_i=0$ which produces the matrix equations
\eqn\supercom{[X^\mu,X^\nu]=0}
or in components,
\eqn\superrel{\eqalign{
X^1_{i,i+a_1} X^2_{i+a_1,i+a_1+a_2} = X^2_{i,i+a_2} X^1_{i+a_2,i+a_2+a_1} \cr
X^2_{i,i+a_2} X^3_{i+a_2,i+a_2+a_3} = X^3_{i,i+a_3} X^2_{i+a_3,i+a_3+a_2} \cr
X^3_{i,i+a_3} X^1_{i+a_3,i+a_3+a_1} = X^1_{i,i+a_1} X^3_{i+a_1,i+a_1+a_3} .
}}
Although na\"{\i}vely there are $3n-3$ conditions, which combined with the
$n-1$ quotients by $U(1)$ would eliminate the (Higgs branch) moduli space,
these conditions are not transverse to one another  and thus the solution to
\superrel\
can be a non-trivial variety, which we denote by $\CN_{\vec a}$.
For example,
it is easy to verify that with no Fayet--Iliopoulos terms,
the Higgs branch $\CN_{\vec a}\gitquot U(1)^{n-1}$ coincides with
$\BC^3/\BZ_n$, e.g. by taking $X^\mu = x^\mu \gamma(g)^{a_\mu}$.

The Fayet--Iliopoulos coefficients
$\zeta_i$ control the metric on moduli space, and
as in \dm\ this means they should be controlled by twisted moduli.
The computation of these couplings
in the appendix of \dm\ can be extended to this case.
The main point is the r\^ole of the Chan--Paton twist $\gamma(g)$ in the
amplitude---a coupling to the sector twisted by $g^k$ corresponds to
a world-sheet amplitude with an insertion of $\gamma(g^k)$ at the
end-point of the cut produced by the closed string twist field.
The couplings between the R-R fields and gauge field strengths on the
branes are then easy to compute, while the Fayet--Iliopoulos terms are the
supersymmetric
partners.
This leads (in the example of the \IIb\ D$1$-brane, with the other D$p$-branes
being analogous) to couplings of the form
\eqn\twistcouple{
\sum_{k=1}^{n-1} \int d^2x\
\tilde C^{(2)}_k \Tr \gamma(g^k)
+ \tilde C^{(0)}_k \Tr \gamma(g^k) \CF
+ \phi_k \Tr \gamma(g^k) \CD
}
where $\CF$ is the  matrix of
gauge field strengths (satisfying the projection
$\gamma(g)\CF\gamma(g)^{-1}=\CF$), and
$\CD$ the corresponding matrix of auxiliary fields.
Since $\gamma(g^k)_{ii}=\omega^{ik}$, the coupling involves a discrete
Fourier transform.
In the $\BZ_3$ example, we have $\phi_2=\phi^*_1$ and
the coefficients of the D-terms become
$$\zeta_i = \omega^i\phi_1 + \omega^{2i} \phi^*_1$$
which satisfy $\sum\zeta_i=0$ and
see both real and imaginary parts of $\phi_1$, in other words,
both $J$ and $B$.

\subsec{``Fractional branes'' and wrapped branes}

As discussed in \egs, world-sheet consistency conditions do not require the
image D-branes to fill out the regular representation of $\Gamma$.
Even if we start with the regular representation, when the images sit
at the fixed point, there are flat directions which separate them in
the non-compact space (the Coulomb branch) \polpro.  Thus if we claim
that the orbifold resolution is geometric, we must find a geometric
interpretation for these states.

Let us consider the world-volume theories associated with the
one-dimensional representations of $\Gamma$.  Specifically, for each $k$,
$1\le k\le n$, we
can consider the representation $\gamma^{(k)}$ defined by
$\gamma^{(k)}(g)_{ij}=\delta_{ik}\delta_{jk}\omega^k$.
The orbifold projection removes the transverse coordinates and
the only surviving matter is the position in the non-compact space, so this
theory describes an object bound to the fixed point.
Its mass and charge can be determined by doing world-sheet computations
of the one-point function of the metric or R-R string vertex operators,
and the leading term comes from a disk diagram with boundary on the
D-brane.  The dependence on the representation $\gamma$ is the factor
$\Tr(\gamma^i)$ in the $i\th$ twisted sector as in \twistcouple.

Since each image couples equally to the metric, in conventions where
a conventional (regular representation) brane has mass $1/g_s$,
a single image has mass $1/n g_s$.
This is exact at the orbifold point, while turning on background fields
can modify the mass, as we discuss shortly.
Similarly, in conventions where the R-R charge of a conventional brane is $1$,
the brane associated with the $k\th$ representation has charge
$\omega^{ik}/n$ for the R-R field in the $i\th$ twisted sector.
This is the same discrete Fourier transform which appears in the gauge
field couplings and it is simpler to think of the $k\th$ representation
as associated with the $k\th$ $U(1)$ in the quiver diagram, and work
with R-R fields in the Fourier transform basis as well.  The scalars
in this basis control Fayet--Iliopoulos parameters; as above the $k\th$ and
$(n{-}k)\th$
scalar control $B$ and $J$ for the $k\th$ two-cycle.
Thus the $k\th$ brane is charged under the R-R fields associated with
the $k\th$ (or $(n{-}k)\th$) two-cycle and four-cycle.  (This discussion is
closely related
to the generalized McKay correspondence which holds for arbitrary finite
subgroups of SU(3) \refs{\itoreid,\reid}.)

The natural interpretation of all this is that the $k\th$ ``fractional
brane'' is actually a
Dirichlet $(p+\ell)$-brane, wrapped around an $\ell$-cycle implicit in the
orbifold theory, to produce a D$p$-brane in the lower dimensional theory
(or a bound state of such branes) \refs{\polpro,\egs}.
Further confirmation of this can be found by considering the supersymmetric
partners of these couplings.
For example, turning on the Fayet--Iliopoulos parameter $\zeta_i$
for the $U(1)$ associated with the brane increases its mass by
$\zeta_i^2/n g_s$, the only term in the D-term potential surviving the
projection.  On the other hand, the brane still has the same fermion
zero modes and corresponds to an object in the same supermultiplet,
so the result must be a mass for the brane consistent with the central
charge formula.

In gauge theory terms, turning on $\zeta_i$ eliminates
the supersymmetric vacua on the Higgs branch; however the
supersymmetry breaking is of a trivial form, a constant shift of
the Hamiltonian.  This must be associated with a central charge
and it would be interesting to make this precise.

These results for the masses are controlled at weak string coupling and
for small blow-up parameter $|\zeta|<<\ap$.  For $|\zeta|\sim\ap$, we would
expect them to be modified by corrections to the D-brane world-volume
Lagrangian analytic in $\zeta/\ap$, defined in principle by
string world-volume computation.  The phase boundary between orbifold
and non-linear sigma model phases \agm\ might find a counterpart here
in the domain of convergence of this expansion.

In the $\BZ_3$ example, the three elementary wrapped branes have
charges $(1,0)$, $(-\half,\sqrt{3}/2)$ and $(-\half,-\sqrt{3}/2)$
under the real and imaginary parts of $\tilde C^{(p)}_1$.
In a more natural charge basis, the charges
of the three states are $(1\ 0)$, $(0\ \half)$ and $(-1\ -\half)$; not having
computed the kinetic term for $\tilde C^{(p)}_1$, the normalization
of these charges is undetermined, so we might as well use this basis.

In \IIa\ theory, we could consider ``fractional'' zero-branes; what
states are these connected to in the
large radius limit?  Zero-branes can come from wrapped D$2$-branes,
wrapped D$4$-branes or bound states of these objects.  These objects
are electrically and magnetically charged (respectively) under the
$U(1)$ associated with the two-cycle.  Evidently the first object is
the wrapped two-brane and the second two are wrapped four-branes;
the electric charge of the wrapped four-branes comes from the expectation
value for $B$ and the coupling $\int C^{(2)} \wedge (F-B)$.
The symmetry between the two wrapped four-branes fits well with the
known value $B=\half$ at the orbifold point \rAGMsd.

The masses $1/n g_s$ of the elementary fractional branes
implicitly determine the K\"ahler modulus
$A=\int B+iJ$ at the orbifold point, as well as its special geometry
partner $\CF'(A)$ determining the masses of magnetically charged states.
For example, the mass spectrum of the $\BC^3/\BZ_3$ case suggests that
the gauge coupling is $\tau = e^{2\pi i/3}$, which
is natural as the R-R gauge field coupling does not depend on $g_s$.

All of these masses can also be determined by mirror symmetry techniques
\refs{\rAGMsd,\rGK}, though not all of these predictions have been made
explicit yet;
for example the masses of wrapped four-branes (currently under investigation).
Completing this comparison, besides confirming these identifications,
would give a microscopic explanation of the masses of wrapped branes.
Symmetries which can only exist in quantum geometry, such as the $\BZ_3$
acting on the wrapped two-branes and four-branes of our example, would
also be given an explicit microscopic picture.

Besides the elementary branes associated with one-dimensional
representations, the general theory of this form can contain
bound states at threshold, which will correspond to wrapped branes
of higher degree.  Such branes are expected to exist at large volume
\witPT\ and it will be interesting to identify these as well.

\subsec{Further corrections to the world-volume Lagrangian}

In the case of $\BC^2/\Gamma$ considered in \dm, supersymmetry determined
the kinetic term in the D-brane world-volume Lagrangian,
preventing other dependence on
the twist fields $\phi_i$.
 (See \dos\ for further discussion, especially
of the case of several D$0$-branes.)
In the present case, this is no longer true.
The configuration space can have a non-trivial
metric, as long as it is asymptotically flat and reduces to the flat metric
for $\phi_i=0$.  In principle, it can be determined at weak string coupling
by explicit world-sheet computation, again along the lines of \dm.

The metric could also get corrections in the string loop expansion.
For general $p$-branes, these could be important even at weak string coupling,
due to world-volume IR effects (i.e., renormalization).
However, for D$0$-branes, the dynamics are governed by a non-singular
quantum mechanical theory of
heavy objects, and as such, IR effects are controlled by the masses of
the world-volume degrees of freedom, which in turn are determined by the
blow-up parameters $\zeta$.  Thus the loop expansion will be
controlled by the parameter $g_s/\zeta^{3/2}$.
For blow-ups which are large compared to the eleven-dimensional Planck
scale but small compared to the string scale, these effects are small,
justifying the classical description.

On the other hand, these corrections are uncontrolled in
the strong coupling limit, and the bare metric must be regarded as
undetermined.
Thus, the definition of M theory on an orbifold advocated in \egs,
as the strong coupling limit of the D$0$-brane theory (which if interpreted
following the proposal of \BFSS\ could give a complete definition of
M theory in this background), is not fully explicit
for $\BC^3/\Gamma$.

This is an important point because the simplest version of the
construction\ifx\answ\bigans\else\ \fi---namely, to use the flat metric for
the D-brane gauge theory
configuration space---appears to disagree with physical expectations.
M theory for low energy processes on manifolds of small curvature
$(R \ll 1/l_{pl}^2)$ reduces to supergravity, and thus one might expect the
low energy configuration space in the problem at hand to have a Ricci-flat
metric.
The classical treatment (which can be justified for D$0$-branes in this
regime) produces the metric as a K\"ahler quotient, which on general grounds
has no reason to be Ricci-flat.  Indeed, Sardo Infirri has argued
that in the example of $\BC^3/\BZ_3$, the relevant metric is {\it not}\/
Ricci-flat \infirri.

One possible resolution of this discrepancy is that the bare metric
might not be flat.\foot{A similar phenomenon has recently been observed in
an analogous context \bss.}
Another is that bound states of large numbers of D$0$-branes see
an effective metric which is Ricci flat.
These issues will be discussed in \dg\ and \dta,
but for present purposes,
when we make statements about M theory, we are assuming that some form
of the D$0$-brane system defined here, in particular
with a non-singular kinetic term, is a correct representation of the
strong coupling limit.  This will allow us to make certain topological and
even geometric statements.

\newsec{Geometrical interpretation}

In this section we shall consider the topological properties
of the D-brane vacuum moduli space. Since sub-string scale spacetime
emerges from this moduli space, we anticipate that the construction
just outlined is geometrically interpretable as $\BC^3/\Gamma$ with
the smoothing of its singularities being controlled by the Fayet--Iliopoulos
parameters. As we reviewed,  previous studies have shown \refs{\agm,\rWP},
though, that
perturbative string theory probes a rich phase structure associated
with an orbifold background. For various choices of the blow-up parameters,
the resulting configuration can be any of a number of birationally
equivalent but topologically distinct smooth Calabi--Yau spaces,
partially resolved orbifolds, Landau--Ginzburg orbifolds, or hybrids
which mix these possibilities.
{}From the perspective of perturbative string theory
all of these distinct backgrounds smoothly
join on to one another as the Fayet--Iliopoulos parameters are varied.
By interpreting the D-brane vacuum moduli space as an effective
spacetime background, we will be able to determine the fate of these
phases at ultra-short distances.

The classical D-brane moduli space, for fixed values of the
Fayet--Iliopoulos parameters, is obtained by imposing the superpotential
\superrel\ and D-term constraints \dterms, and quotienting by the residual
$U(1)^{n-1}$ gauge symmetry group.  Our analysis of this moduli space will
be based on the observation---also made by Sardo Infirri \infirritwo---that
this space can also be described using toric geometry, or equivalently,
described as the
classical moduli space for an abelian gauged linear sigma model.  The toric
description has the pleasant feature of treating the
superpotential and D-term constraints in a more-or-less uniform manner.

\subsec{A toric presentation}

To make the translation into a toric description, we begin by
solving \superrel\ on a dense open subset.
Let us choose our generator of $\Gamma = \BZ_n$ to have $a_3 = -1$
and define $a=a_1$ and $b=a_2$. Defining
$x_i=X^1_{i,i+a_1}$, $y_i=X^2_{i,i+a_2}$ and $z_i=X^3_{i,i+a_3}$, we note that
we can use  \superrel\ to solve for  $2n-2$ of the variables
in terms of the remaining set of $n+2$. In particular, we can solve for
 $x_k$ in terms of $x_{k-1}$ and the $z_i$
as
\eqn\itrate{ x_k = {z_k\over z_{a+k}} x_{k-1}.
}
Iterating this produces
\eqn\xsol{\eqalign{
x_k &= {z_1\ldots z_k\over z_{a+1}\ldots z_{a+k}} x_{0} \cr
&= {z_1 \ldots z_k\ z_1 \ldots z_a\over z_1 \ldots z_{a+k}} x_{0}.
}}
with all indices taken modulo $n$.
Similarly,
\eqn\ysol{
y_k = {z_1 \ldots z_k\ z_1 \ldots z_b\over z_1 \ldots z_{b+k}} y_{0}.
}

As a consequence, we see that $\CN_{\vec a}\subset \IC^{3n}$ contains a
copy of $(\IC^*)^{n+2}$ as an open subset, with an action of
$(\IC^*)^{n+2}$ on $\CN_{\vec a}$ which restricts to the usual action by
multiplication on that subset.  This is one of the standard
definitions of a toric variety
(see \refs{\agm,\oda,\fulton} for an introduction).
In fact, since $\CN_{\vec a}$ is a subvariety of $\BC^{3n}$ defined by
monomial
relations, it is an ``affine toric variety,'' which is somewhat simpler
than the general case.

Toric geometry describes a space by the well-defined algebraic functions it
supports.  Any monomial in the coordinates $x_i$, $y_i$ and $z_i$ defines
such a function on $\CN_{\vec a}$, but they are not all independent---as we
have seen, we can use \xsol\ and \ysol\ to express $2n-2$ of the basic
coordinate functions in terms of $x_0$, $y_0$ and $z_i$.
Changing notation temporarily, we let
$(v_{-2},v_{-1},v_0,\dots,v_{n-1})=(x_0,y_0,z_0,\dots,z_{n-1})$ and
$(w_0,...,w_{3n-1}) =
(x_0,...,x_{n-1},y_0,...,y_{n-1},z_0,...,z_{n-1})$.
Then we can write
$w_i=\prod v_j^{m_{ij}}$
and regard each $\vec{m}_i\in M=\BZ^{n+2}$ as a vector of exponents.
A general monomial will now be associated with a point in $M$
given by a sum of these basis vectors
with non-negative integer coefficients.
The set of all such points will form a cone $M_+\subset M$.

Conversely, a basis for the algebra of
functions on $\CN_{\vec a}$ determines the
space $\CN_{\vec a}$.
The cone $M_+$ determines such an algebra and basis---each point in the
cone gives us a basis vector,
and the multiplication
law is induced from addition in $\BZ^{n+2}$.

Since we can reconstruct $\CN_{\vec a}$ from the cone $M_+$,
we can use any description of the cone to define the space.
Another description  is in terms of a set of hyperplanes which
bound the cone, i.e., vectors $\vec n$ for which
\eqn\posi{\vec n \cdot \vec m \ge 0.}
These vectors are points in the dual lattice $N\cong \Hom(M,\BZ)$.
Clearly the sum of two solutions $\vec n$
of \posi\ will also satisfy \posi\ for
all $M_+$, so this condition defines a dual cone $N_+\subset N$.
Given $N_+$, the original cone $M_+$ can be reconstructed, so
the space $\CN_{\vec a}$ can also be specified by a choice of cone $N_+$.

The advantage of the cone $N_+$ is that it allows us to give an alternate
description of the toric variety $\CN_{\vec a}$ as a quotient.
There are two approaches
to presenting a toric variety as a quotient
\refs{\rDelzant,\rAudin}---it can be thought of as a holomorphic quotient,
or (if K\"ahler) as a symplectic quotient.
(The relationship between the two is discussed in some
detail in \refs{\rAG,\mp} to which the reader can refer for more details.)
In a nutshell, a toric variety of dimension $k$ can be expressed in the form
$(\IC^q {-} F_{\Delta}) / (\IC^*)^{q-k}$ for some set $F_{\Delta}$ and
some $(\IC^*)^{q-k}$ action in $\IC^q$. The latter action can be carried out
as written (the
holomorphic quotient) or (in the K\"ahler case) it can be carried out in
two steps, an $(\IR_+)^{q-k}$ action and
a $U(1)^{q-k}$ action (the  symplectic quotient). The resulting  quotient
space depends on the precise form of the gauge fixing determined
by the  $(\IR_+)^{q-k}$ action, the so-called moment map.
The distinct possibilities, in fact, are the phases of \refs{\agm,\rWP}.
In the holomorphic approach,
these phases are distinguished by being associated to different triangulations
of certain point sets which in turn determine different point sets
$F_{\Delta}$ to be removed from the initial parent space.
If all possibilities are considered---that is, all triangulations
and all gauge fixings---then all  phases are accessed. This is the case for
perturbative closed string
theory. If, however, only a subset of gauge fixings are relevant
for the physical model, then  only some of the  phases are physically
realized. This turns out to be the case for D-branes.

To implement this quotient description, we introduce
homogeneous coordinates $p_i, i = 0,\dots, q-1$  in the sense of \rCox,
associated to a set $\cascade$ consisting of $q$ points in the lattice $N$.
Taking integer linear combinations of these points defines a natural map
\eqn\enatmap{
T:\BZ^{\cascade}\to N,}
which we assume to be surjective.  The transpose of the kernel of $T$ is
then a $(q-k)\times k$ ``charge matrix'' $Q$, which specifies an action
of $U(1)^{q-k}$ (or $(\IC^*)^{q-k}$) on $\IC^q$.  A triangulation $\Delta$
of the convex hull of $\cascade$ then determines---through a specific
combinatorial procedure---a subset $F_\Delta\subset \IC^q$ such that the
toric variety takes the form $(\IC^q{-}F_\Delta)/(\IC^*)^{q-k}$.  (See
\agm\ for the combinatorial details of this step.)  In the case of the
toric variety $\CN_{\vec a}$,
the point set $\cascade$ consists of the generators of the cone $N_+$.

For our present purposes, we also need to see how the action of $(\IC^*)^k$
on the toric variety (which was part of its definition) is represented in
these terms.  Since $\IC^q{-}F_\Delta$ always contains $(\IC^*)^q$, our
toric variety $(\IC^q{-}F_\Delta)/(\IC^*)^{q-k}$
contains $(\IC^*)^q/(\IC^*)^{q-k}\cong (\IC^*)^k$ as an open subset.
That group will act on the toric variety through an
action of $(\IC^*)^k$  on $\IC^q$, provided that the projection
of that action to $(\IC^*)^q/(\IC^*)^{q-k}\cong (\IC^*)^k$ gives the
identity map.  Such an action is specified by a $k\times q$ charge matrix
$U$, and the condition on the projection is that
\eqn\econdp{
T\cdot {}^tU = {\rm Id}_k,}
where $T$ is the $k\times q$ matrix \enatmap\ specifying the toric variety.
We choose such a matrix $U$.

{}From the point of view of Witten's gauged linear sigma model, this
mathematical construction can be physically realized by  interpreting
the  $p_i$ as
chiral superfields interacting in a $U(1)^{q-k}$ two-dimensional
gauged linear sigma model. Inclusion of the Fayet--Iliopoulos D-terms
with coefficients $\xi_j, j= 1,...,q-k$ provides the degrees of freedom
which are the counterparts to the possible triangulations of the
$p_i$. It is by now well known that the formulation of a toric variety using
the holomorphic quotient of the last paragraph is equivalent to the Witten's
physical incarnation since the latter realizes the same toric variety
as a symplectic quotient $\IC^q\gitquot U(1)^{q-k}$. So, given a D-brane
configuration, we can follow the procedure of this section to generate
mathematical toric data which can then be re-translated into the physics of
a gauged linear sigma model.

{}From a topological point of view, if the $\xi_j$ are completely generic,
then all possible phases of the linear sigma model are physically realized,
corresponding to all $c_1 = 0$ birational transformations on the toric
variety $V$. That is, backgrounds corresponding to all possible triangulations
of the point set $\cascade$ are realized. If, on the contrary, only a
restricted
set of values for the $\xi_j$ are physical, then only a subset of the possible
phases will be probed. This is precisely what happens in constructing
the D-brane moduli space. Namely, as we shall see below in the context
of two examples, the D-brane vacuum
moduli space on $\IC^3/\BZ_n$ gives precisely the standard toric point set
$\cascade$ for the blow-up of $\IC^3/\BZ_n$. However, the  $\xi_i$
parameters, determined by the $\zeta_j$ from the last section, are
{\it not}\/ generic and therefore only part of the phase space of
the blown-up orbifold is probed by D-branes. Although we do not
have a general proof, in the examples we have studied (and in examples
studied by Sardo Infirri \infirritwo) the phases
which are eliminated are {\it non-geometric}\/ phases reviewed earlier.
This  dovetails nicely with the result of Witten
in \witPT\ in which he found that the non-geometric phases are the ones
which get squeezed out when passing to strongly coupled type IIA string
theory---that is, M-theory. It is in this limit that D$0$-branes dominate
low energy physics, and hence the fact that their vacuum moduli
space---which we interpret as an effective spacetime background---does not
see the non-geometric phases is in agreement with Witten's result.

To see this explicitly, we follow the procedure outlined above. Namely,
we express the solution to the  superpotential constraint in terms of
a cone $M_+$. We take its dual to get a cone $N_+$.
The $q$ generators of $N_+$ determine a map $T:\BZ^{\cascade}\to N$, and
the transpose of the kernel of $T$ determines a $(q-n-2)\times q$ charge
matrix $Q$.  In order to obtain $\CN_{\vec a}$ as the symplectic quotient
$\IC^q\gitquot U(1)^{q-n-2}$, {\it we must set the associated D-terms
$\xi_1$, \dots, $\xi_{q-n-2}$ to zero.}  To describe the action of the
$U(1)^{n-1}$ charges in the twisted sectors on $\CN_{\vec a}$,
we represent the action of $U(1)^{n-1}$ on $\IC^{n+2}$ by an
$(n-1)\times (n+2)$ charge matrix $V$; the product $VU$ will then give an
$(n-1)\times q$ charge matrix for the action of $U(1)^{n-1}$ on $\IC^q$.
(This charge matrix depends on the choice of a matrix $U$ satisfying
\econdp, but different choices will only differ by charges from the matrix
$Q$.) Non-zero values of the corresponding D-terms
$\xi_i=\zeta_{i-(q-n-2)}$ {\it are}\/ allowed.

The full set of charges is now given by a $(q-3)\times q$ charge matrix
$\widetilde{Q}$, the concatenation of $Q$ and $VU$.  The cokernel of its
transpose gives toric data for the D-brane vacuum moduli space, in the form
of a map $\widetilde{T}:\BZ^{\cascade}\to \BZ^3$.
We claim that this data
is nothing but the usual toric data for resolving the quotient space
$\IC^3/\BZ_n$ (supplemented by some extra auxiliary fields as well as some
extra $U(1)$'s which can be used to eliminate them)
showing an explicit realization of the D-brane moduli space aligning with
the underlying spacetime structure. Second, we also claim that the
non-genericity
of the Fayet--Iliopoulos parameters is such that the non-geometric
phases of this desingularization are not physically realized.

We start with the case $n = 3$. Following the procedure outlined,
the cone $M_+$ is generated by the rows in the following:
\eqn\coneM{\matrix{
&&x_0&y_0&z_0&z_1&z_2 \cr
x_0&&1&0&0&0&0 \cr
x_1&&1&0&-1&1&0 \cr
x_2&&1&0&-1&0&1 \cr
y_0&&0&1&0&0&0 \cr
y_1&&0&1&-1&1&0 \cr
y_2&&0&1&-1&0&1 \cr
z_0&&0&0&1&0&0 \cr
z_1&&0&0&0&1&0 \cr
z_2&&0&0&0&0&1, }
}
and the charges of the fields $(x_0,y_0,z_0,\dots,z_2)$ under the twisted
sector gauge group $U(1)^2$ (with D-terms $\zeta_1$, $\zeta_2$, as in
\dterms) are given by
\eqn\etwistchg{
V=\pmatrix{
0 & 0 & 0 & -1 & 1 \cr
1 & 1 & 1 & 0 & -1
}.}
The dual cone $N_+$ is generated by the columns of
\eqn\coneN{
T=\pmatrix{
1 & 0 & 0 & 1 & 0 & 0 \cr
0 & 1 & 0 & 1 & 0 & 0 \cr
0 & 0 & 1 & 1 & 0 & 0 \cr
0 & 0 & 1 & 0 & 1 & 0 \cr
0 & 0 & 1 & 0 & 0 & 1}
}
(which correspond to homogeneous coordinates $p_0, \dots, p_5$).  The
columns of $T$ were found as solutions of the equations \posi, with the
monomial corresponding to the hyperplane $\vec n$ being
$\prod X_i^{\vec n\cdot \vec m_i}$.

The transpose of the kernel of this matrix is
\eqn\echargmat{
Q=\pmatrix{
1 & 1 & 1 & -1 & -1 & -1
}.}
We choose $U$ satisfying \econdp\ to be
\eqn\eU{
U=\pmatrix{
0 & -1 & -1 & 1 & 1 & 1 \cr
0 & 1 & 0 & 0 & 0 & 0 \cr
0 & 0 & 1 & 0 & -1 & -1 \cr
0 & 0 & 0 & 0 & 1 & 0 \cr
0 & 0 & 0 & 0 & 0 & 1
},}
so that the full charge matrix $\widetilde{Q}$ obtained by concatenating
$Q$ and $VU$ is
\eqn\efull{
\widetilde{Q}=\pmatrix{
1 & 1 & 1 & -1 & -1 & -1 \cr
0 & 0 & 0 & 0 & -1 & 1 \cr
0 & 0 & 0 & 1 & 0 & -1
}.}

The cokernel of the transpose of $\widetilde{Q}$ (which can be calculated
as the transpose of the kernel) is the matrix
\eqn\ecoktr{
\pmatrix{
-1 & 1 & 0 & 0 & 0 & 0 \cr
-1 & 0 & 1 & 0 & 0 & 0 \cr
 3 & 0 & 0 & 1 & 1 & 1
}.}
This matrix has three identical columns, which means that the point set
$\widetilde{\cascade}$ determined by those columns actually only contains
four distinct elements.  If we eliminate the redundant variables and
$U(1)$'s, we obtain a smaller matrix
\eqn\ecoktrs{
\widetilde{T}=
\pmatrix{
-1 & 1 & 0 & 0 \cr
-1 & 0 & 1 & 0 \cr
 3 & 0 & 0 & 1
}}
specifying the toric data for the D-brane moduli space $\CN_{\vec
a}\gitquot U(1)^{n-1}$.

We recognize this as the toric data for resolving $\IC^3/\BZ_3$.  (The
charge matrix determined by $\widetilde{T}$ is $(1\quad1\quad1\quad-3)$.)
Triangulating on all points yields the blow-up ${\cal O}(-3) \rightarrow
\BP^2$. Leaving out the central point, in the context of type II string
theory, would take us to the Landau--Ginsburg phase, which is geometrically
associated with the blown-down singular variety. In Witten's gauged linear
sigma
model,
these phases arise from studying the bosonic potential
\eqn\epotential{U = (|u_1|^2 + |u_2|^2 + |u_3|^2 - 3|p|^2 - \xi)^2}
for $\xi$ positive and for $\xi$ negative. In the former case, vanishing
of the potential is consistent with $p = 0$, which directly gives
the exceptional divisor, $\BP^2$ with size controlled by $\xi$.
When $\xi$ is negative, though, $p$ cannot vanish, corresponding to the
exceptional divisor being blown down.

In our case, we have precisely the same toric point set (the columns of
\ecoktrs)
but the Fayet--Iliopoulos parameters $\zeta_i$ are only associated with
the last two $U(1)$ quotients in \efull\ since the first
actually originates from a superpotential constraint in the original
D-brane action. As $\zeta_1$ and $\zeta_2$ take on generic
values, it is not hard to see that the corresponding value of
$\xi$ is non-negative, thereby only accessing the blown-up phase.
To see this  note that for any non-zero values
of the $\zeta_i$, the second two D-terms imply that
two out of the last three homogeneous variables $p_3,p_4,p_5$ are
nonzero. Calling the
one that can potentially  vanish $\tilde p$, we can solve the latter
two D-term constraints in terms of $\tilde p$ and therefore
rewrite the first D-term as
\eqn\epotentialD{ (|p_0|^2 + |p_1|^2 + |p_2|^2 - 3|\tilde p|^2 -
f(\zeta_1,\zeta_2))^2.}
We claim that $f(\zeta_1,\zeta_2)>0.$  In fact, if we write
the charge matrix again with an extra column including the D-terms
\eqn\erewrite{
\pmatrix{
1 & 1 & 1 & -1 & -1 & -1 & 0 \cr
0 & 0 & 0 & 0 & -1 & 1 & {{ \zeta}_{1}} \cr
0 & 0 & 0 & 1 & 0 & -1 & {{ \zeta}_{2}}
},}
and then modify that augmented matrix by invertible row operations to
\eqn\erowmod{
\pmatrix{
1 & 1 & 1 & 0 & 0 & -3 &  - {{ \zeta}_{1}} + {{ \zeta}_{2}} \cr
0 & 0 & 0 & 0 & 1 & -1 &  - {{ \zeta}_{1}} \cr
0 & 0 & 0 & 1 & 0 & -1 & {{ \zeta}_{2}}
},}
we see that when $\zeta_1<0$ and $\zeta_2>0$, we have $p_4\ne0$ and
$p_3\ne0$, and also $f(\zeta_1,\zeta_2)=-\zeta_1+\zeta_2>0$.  This verifies
our claim in one of three phase regions, illustrated in figure 2.  Similar
verifications can be made in the other two regions, using the alternate forms
\eqn\evariants{
\pmatrix{
1 & 1 & 1 & -3 & 0 & 0 &  - {{ \zeta}_{1}} - 2\,{{ \zeta}_{2}} \cr
0 & 0 & 0 & -1 & 1 & 0 &  - {{ \zeta}_{1}} - {{ \zeta}_{2}} \cr
0 & 0 & 0 & -1 & 0 & 1 &  - {{ \zeta}_{2}}
}\hbox{\ and\ }\pmatrix{
1 & 1 & 1 & 0 & -3 & 0 & 2\,{{ \zeta}_{1}} + {{ \zeta}_{2}} \cr
0 & 0 & 0 & 0 & -1 & 1 & {{ \zeta}_{1}} \cr
0 & 0 & 0 & 1 & -1 & 0 & {{ \zeta}_{1}} + {{ \zeta}_{2}}
}}
of \erewrite.

\ifig\figtwo{Phases in the $(\zeta_1,\zeta_2)$ plane.}
{\epsfxsize3.0in
\epsfbox{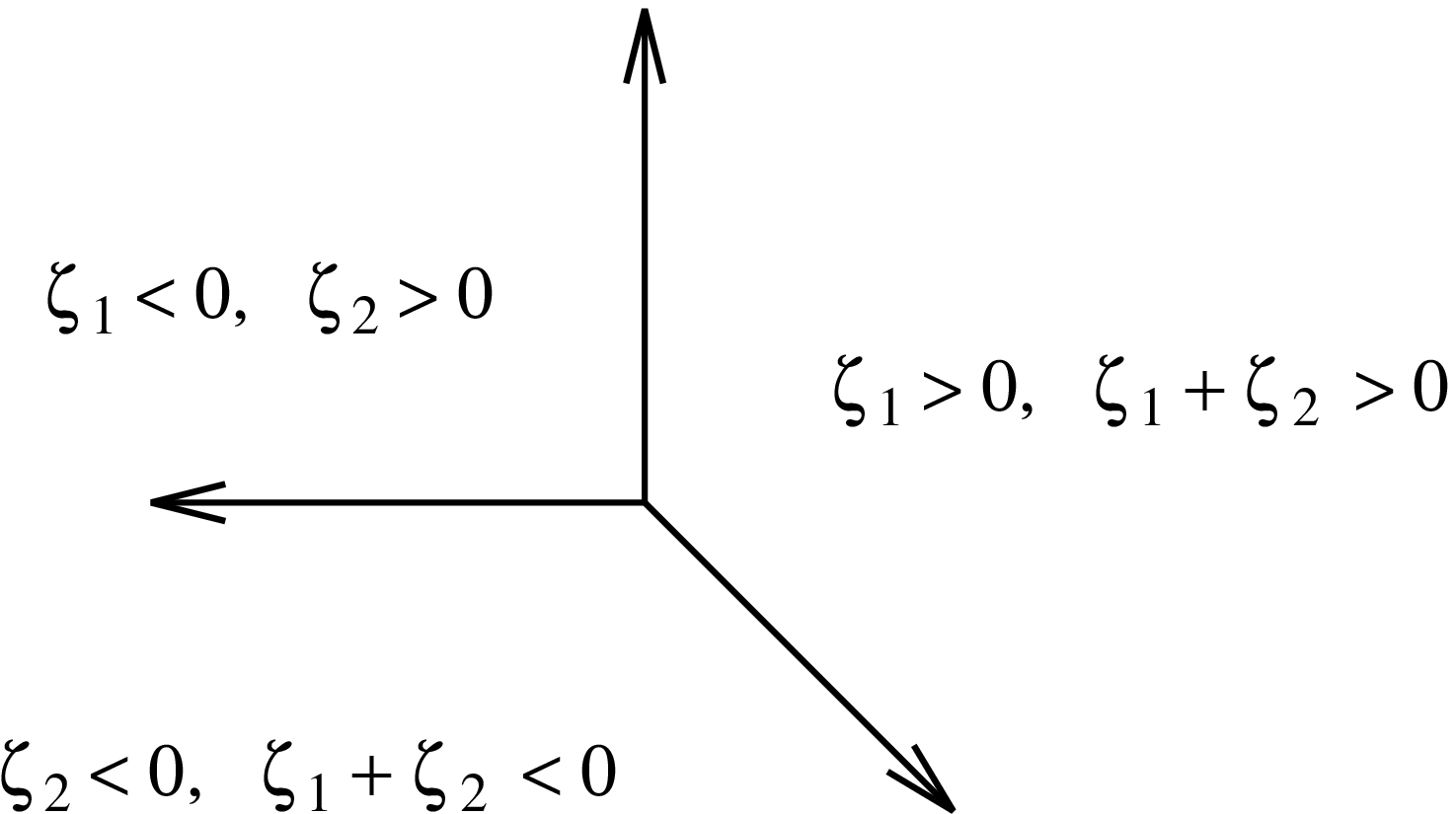}}

The case of $n = 5$ is similar, but technically more involved. Taking
the action to be $(Z^1,Z^2,Z^3) \rightarrow (\omega^3 Z^1, \omega^3 Z^2,
\omega^{-1} Z^3)$,
we find that the cone $M_+$ is generated by the rows in the following:

\eqn\coneMFive{\matrix{
&&x_0&y_0&z_0&z_1&z_2&z_3&z_4 \cr
x_0&&1&0&0&0&0&0&0 \cr
x_1&&1&0&0&1&0&0&-1 \cr
x_2&&1&0&-1&1&1&0&-1 \cr
x_3&&1&0&-1&0&1&1&-1\cr
x_4&&1&0&-1&0&0&1&0\cr
y_0&&0&1&0&0&0&0&0 \cr
y_1&&0&1&0&1&0&0&-1 \cr
y_2&&0&1&-1&1&1&0&-1\cr
y_3&&0&1&-1&0&1&1&-1\cr
y_4&&0&1&-1&0&0&1&0 \cr
z_0&&0&0&1&0&0&0&0 \cr
z_1&&0&0&0&1&0&0&0 \cr
z_2&&0&0&0&0&1&0&0 \cr
z_3&&0&0&0&0&0&1&0\cr
z_4&&0&0&0&0&0&0&1, }
}
and the charges of the fields $(x_0,y_0,z_0,\dots,z_4)$ under the twisted
sector gauge group $U(1)^4$ (with D-terms $\zeta_1$, \dots, $\zeta_4$, as in
\dterms) are given by
\eqn\etwistchgf{
V=\pmatrix{
0 & 0 & 0 & -1 & 1 & 0 & 0 \cr
0 & 0 & 0 & 0 & -1 & 1 & 0 \cr
1 & 1 & 0 & 0 & 0 & -1 & 1 \cr
0 & 0 & 1 & 0 & 0 & 0 & -1
}.}
The dual cone $N_+$ is generated by the columns of
\eqn\coneNf{
T=\pmatrix{
1 & 0 & 0 & 1 & 0 & 0 & 0 & 0 & 1 & 0 & 0 & 0 & 1 \cr
0 & 1 & 0 & 1 & 0 & 0 & 0 & 0 & 1 & 0 & 0 & 0 & 1 \cr
0 & 0 & 1 & 1 & 0 & 0 & 1 & 1 & 1 & 0 & 0 & 0 & 0 \cr
0 & 0 & 1 & 0 & 1 & 1 & 1 & 0 & 0 & 1 & 0 & 0 & 0 \cr
0 & 0 & 1 & 1 & 1 & 0 & 0 & 1 & 0 & 0 & 1 & 0 & 0 \cr
0 & 0 & 1 & 0 & 0 & 1 & 1 & 1 & 0 & 0 & 0 & 1 & 0 \cr
0 & 0 & 1 & 1 & 1 & 1 & 0 & 0 & 0 & 0 & 0 & 0 & 1
}}
(which correspond to homogeneous coordinates $p_0, \dots, p_{12}$).
We choose $U$ satisfying \econdp\ to be
\eqn\eUf{
U=\pmatrix{
1 & 0 & 0 & 0 & 0 & 0 & 0 & 0 & 0 & 0 & 0 & 0 & 0 \cr
0 & 1 & 0 & 0 & 0 & 0 & 0 & 0 & 0 & 0 & 0 & 0 & 0 \cr
-1 & -1 & 0 & 0 & 0 & 0 & 0 & 0 & 1 & 0 & 0 & 0 & 0 \cr
0 & 0 & 0 & 0 & 0 & 0 & 0 & 0 & 0 & 1 & 0 & 0 & 0 \cr
0 & 0 & 0 & 0 & 0 & 0 & 0 & 0 & 0 & 0 & 1 & 0 & 0 \cr
0 & 0 & 0 & 0 & 0 & 0 & 0 & 0 & 0 & 0 & 0 & 1 & 0 \cr
-1 & -1 & 0 & 0 & 0 & 0 & 0 & 0 & 0 & 0 & 0 & 0 & 1
},}
so that the full charge matrix $\widetilde{Q}$ obtained by concatenating
$Q={}^t(\ker T)$ and $VU$ (and including an extra column for the D-terms)
is
\eqn\efullf{
\widetilde{Q}=\pmatrix{
1 & 1 & 0 & 2 & -1 & -1 & 2 & -1 & -3 & 0 & 0 & 0 & 0 & 0 \cr
0 & 0 & 1 & -1 & 0 & 0 & -1 & 0 & 1 & 0 & 0 & 0 & 0 & 0 \cr
0 & 0 & 0 & 0 & 0 & 0 & -1 & 1 & 0 & 1 & -1 & 0 & 0 & 0 \cr
0 & 0 & 0 & 0 & -1 & 1 & 0 & 0 & 0 & 0 & 1 & -1 & 0 & 0 \cr
0 & 0 & 0 & 1 & 0 & 0 & 0 & -1 & 0 & 0 & 0 & 1 & -1 & 0 \cr
0 & 0 & 0 & 0 & 0 & -1 & 1 & 0 & -1 & 0 & 0 & 0 & 1 & 0 \cr
0 & 0 & 0 & 0 & 0 & 0 & 0 & 0 & 0 & -1 & 1 & 0 & 0 & {{ \zeta}_{1
}} \cr
0 & 0 & 0 & 0 & 0 & 0 & 0 & 0 & 0 & 0 & -1 & 1 & 0 & {{ \zeta}_{2
}} \cr
0 & 0 & 0 & 0 & 0 & 0 & 0 & 0 & 0 & 0 & 0 & -1 & 1 & {{ \zeta}_{3
}} \cr
0 & 0 & 0 & 0 & 0 & 0 & 0 & 0 & 1 & 0 & 0 & 0 & -1 & {{ \zeta}_{4
}}
}.}

The cokernel of the transpose of $\widetilde{Q}$ (which can be calculated
as the transpose of the kernel) is the matrix
\eqn\ecoktrf{
\pmatrix{
-1 & 1 & 0 & 0 & 0 & 0 & 0 & 0 & 0 & 0 & 0 & 0 & 0 \cr
-3 & 0 & 1 & 0 & 0 & 0 & 0 & 0 & -1 & -1 & -1 & -1 & -1 \cr
5 & 0 & 0 & 1 & 1 & 1 & 1 & 1 & 2 & 2 & 2 & 2 & 2
}.}
This matrix has two sets of identical columns (with five in each set),
which means that the point set
$\widetilde{\cascade}$ determined by the columns actually only contains
five distinct elements.  If we eliminate the redundant variables and
$U(1)$'s, we obtain a smaller matrix
\eqn\ecoktrsf{
\widetilde{T}=
\pmatrix{
-1 & 1 & 0 & 0 &  0 \cr
-3 & 0 & 1 & 0 & -1 \cr
 5 & 0 & 0 & 1 &  2
}}
specifying the toric data for the D-brane moduli space $\CN_{\vec
a}\gitquot U(1)^{n-1}$.
The corresponding charge matrix ${}^t(\ker\widetilde{T})$ is
\eqn\ewTchf{
\pmatrix{
1 & 1 & 0 &  1 & -3 \cr
0 & 0 & 1 & -2 &  1
}.}

We recognize this toric data in $\IR^3$ as that for $\IC^3/\BZ_5$.
{}From the point of view of a symplectic quotient or the gauged linear sigma
model,
this corresponds to five chiral fields with bosonic potential
\eqn\ebosonic{U = (|u_1|^2 + |u_2|^2 + |u_4|^2 - 3|p|^2 - \xi_1)^2 +
 (|u_3|^2 -2 |u_4|^2 +|p|^2 - \xi_2)^2 .}
This linear sigma model has four phases, one of which is the (single)
smooth geometrical resolution (represented by $\xi_1>0$, $\xi_2>0$), two of
which are orbifold phases (in which
one or other of the exceptional divisors has been blown-down) and a
Landau--Ginsburg phase (where this terminology is more apt for models
with a superpotential, but we retain the common labeling scheme).
As before, if we had completely generic Fayet--Iliopoulos terms associated
with each row of the full $U(1)^{10}$ charge matrix \efullf, then the D-brane
moduli space would probe all four of these phases. But, since we only
have non-zero $\zeta_j$ associated with the $U(1)^4$ twisted sector
gauge group, this is not the case. Rather, the three ``non-geometrical''
phases which disappear in the strongly coupled type IIA string are not
realized as the $\zeta$'s take on all possible values: only the geometric
phase is probed by the D-branes.

We will check this statement explicitly only in one phase of the model.  By
invertible row operations, we can modify \efullf\ to
\eqn\efullfv{
\pmatrix{
1 & 1 & 0 & 0 & 0 & 0 & 1 & 0 & 0 & -3 & 0 & 0 & 0 & 2\,{{ \zeta}
_{1}} + 2\,{{ \zeta}_{2}} + {{ \zeta}_{3}} + {{ \zeta}_{4}} \cr
0 & 0 & 1 & 0 & 0 & 0 & -2 & 0 & 0 & 1 & 0 & 0 & 0 &  - {{ \zeta}
_{2}} - {{ \zeta}_{4}} \cr
0 & 0 & 0 & 0 & 0 & 0 & -1 & 1 & 0 & 0 & 0 & 0 & 0 & {{ \zeta}_{1
}} \cr
0 & 0 & 0 & 0 & 1 & 0 & -1 & 0 & 0 & 0 & 0 & 0 & 0 &  - {{ \zeta}
_{2}} - {{ \zeta}_{4}} \cr
0 & 0 & 0 & 1 & 0 & 0 & -1 & 0 & 0 & 0 & 0 & 0 & 0 & {{ \zeta}_{1
}} + {{ \zeta}_{3}} \cr
0 & 0 & 0 & 0 & 0 & 1 & -1 & 0 & 0 & 0 & 0 & 0 & 0 &  - {{ \zeta}
_{4}} \cr
0 & 0 & 0 & 0 & 0 & 0 & 0 & 0 & 0 & -1 & 1 & 0 & 0 & {{ \zeta}_{1
}} \cr
0 & 0 & 0 & 0 & 0 & 0 & 0 & 0 & 0 & -1 & 0 & 1 & 0 & {{ \zeta}_{1
}} + {{ \zeta}_{2}} \cr
0 & 0 & 0 & 0 & 0 & 0 & 0 & 0 & 0 & -1 & 0 & 0 & 1 & {{ \zeta}_{1
}} + {{ \zeta}_{2}} + {{ \zeta}_{3}} \cr
0 & 0 & 0 & 0 & 0 & 0 & 0 & 0 & 1 & -1 & 0 & 0 & 0 & {{ \zeta}_{1
}} + {{ \zeta}_{2}} + {{ \zeta}_{3}} + {{ \zeta}_{4}}
}.}
{}From this form it becomes clear that if the $\zeta_i$'s satisfy
\eqn\econdz{\eqalign{
\zeta_1>0, \quad&-\zeta_2-\zeta_4>0,\quad\zeta_1+\zeta_3>0, \quad-\zeta_4>0,\cr
\zeta_1+\zeta_2>0, \quad&\zeta_1+\zeta_2+\zeta_3>0,\quad
\zeta_1+\zeta_2+\zeta_3+\zeta_4>0,}}
then none of $p_3$, $p_4$, $p_5$, $p_7$, $p_8$, $p_{10}$, $p_{11}$, or
$p_{12}$ can vanish; moreover, $\xi_1=2\,{{ \zeta}
_{1}} + 2\,{{ \zeta}_{2}} + {{ \zeta}_{3}} + {{ \zeta}_{4}}$ and
$\xi_2= - {{ \zeta}
_{2}} - {{ \zeta}_{4}}$ are both positive under these conditions, so we end
up in the geometric phase.

\newsec{Conclusions}

In this paper we have discussed the topological  properties of
D-brane vacuum moduli space in the context of abelian orbifolds
 $\IC^3/\Gamma$ for $\Gamma\subset SU(3)$ and their smooth resolutions.
 We have found that these moduli spaces emerge as
K\"ahler quotients with moment maps determined by  Fayet--Iliopoulos
parameters. Examination of these quotients shows that they are birational
to the singular space $\IC^3/\Gamma$ with the Fayet--Iliopoulos
parameters controlling the size of the blow-up to a smooth space.
This alignment of the vacuum moduli space and the ambient background space
provides another strong piece of evidence that spatial backgrounds at
ultrashort distances can be thought of as a derivative concept emerging
from the fundamental notion of a vacuum moduli space.

The classical vacuum
moduli space coincides with the moduli space of another theory,
a gauged linear sigma model with no superpotential,
motivated by considerations of toric geometry.\foot{It
would be interesting to relate the D-brane worldvolume theory to this
other theory more directly, perhaps by means of a duality transformation in
which the massive states associated to ``fractional branes'' play a
role.}
Hence, the moduli space can {\it a priori}\/ have a rich phase structure.
However, non-genericity of the
Fayet--Iliopoulos parameters appears to ensure that the non-geometric
phases are not part of the physical vacuum D-brane moduli space.

The neighborhood of an orbifold point can contain other transitions.
For example, the $\BC^3/\BZ_{11}$ orbifold has multiple resolutions,
connected by flops.  This D-brane linear sigma model will
contain an explicit geometric realization of the flop, and it will be
interesting to work this out.

It is worth noting that even for $\Gamma\not\subset SU(3)$,
the D-brane theory defined as a quotient of $\CN=4$ super-Yang--Mills
theory is a sensible
non-supersymmetric gauge theory, whose classical moduli space
(on the Higgs branch) is an ALE space asymptotic to $\BC^3/\Gamma$,
also studied in \refs{\infirri,\infirritwo}.  These are non-supersymmetric
theories with
special matter content, and it would be interesting to see if this
moduli space survives in the quantum theory.

\bigskip
\centerline{{\bf Acknowledgments}}\nobreak

We thank the members of the Rutgers high energy theory group for
discussions; B.R.G. and D.R.M. also thank the Rutgers group
for hospitality during various stages of this project.
The work of M.R.D. is supported by DOE grant DE-FG02-96ER40959.
The work of B.R.G. is supported by a National Young Investigator Award and
the Alfred P. Sloan Foundation.
The work of D.R.M. is supported in
part by the Harmon Duncombe Foundation and by NSF grants DMS-9401447
and DMS-9627351.

\listrefs
\end